\documentclass[11pt,prd,superscriptaddress,showpacs,amsmath,amssymb]{revtex4-1}

\usepackage{amsmath}
\usepackage{natbib}
\usepackage{epsf}
\usepackage{slashed}
\usepackage{ulem}
\usepackage{cancel}
\usepackage{color,bm}
\usepackage{graphicx}
\usepackage{amsfonts}
\usepackage{amssymb}
\usepackage{rotating}
\usepackage{booktabs}
\usepackage{xcolor}
\usepackage{soul}
\usepackage{upgreek}
\usepackage{amssymb}
\usepackage{enumerate}
\usepackage{appendix}
\usepackage[caption=false]{subfig}

\definecolor{airforceviolet}{rgb}{0.36, 0.54, 0.66}
\definecolor{steelviolet}{rgb}{0.27, 0.51, 0.71}
\definecolor{amber}{rgb}{1.0, 0.49, 0.0}
\definecolor{mydarkgreen}{RGB}{0,100,0}

\usepackage{changes}
\definechangesauthor[name={HaoSun}, color=purple]{SH}

\usepackage{hyperref}
\hypersetup{
    colorlinks,
    linkcolor={red},
    citecolor={cyan},
    urlcolor={blue!80!black}
}

\definecolor{airforceviolet}{rgb}{0.36, 0.54, 0.66}
\definecolor{steelviolet}{rgb}{0.27, 0.51, 0.71}
\definecolor{amber}{rgb}{1.0, 0.49, 0.0}

\def\red#1{\textcolor{red}{#1}}

\def\comment#1{}

\usepackage{graphicx}

\usepackage[utf8]{inputenc}
\usepackage{textcomp}
\DeclareUnicodeCharacter{2212}{\textminus}
\begin{document}
\title{Searching for exclusive leptoquarks with the Nambu-Jona-Lasinio composite model at the LHC and HL-LHC}
\date{\today}
\author{\textsc{S.~Ajmal}}
\affiliation{Dipartimento di Fisica e Geologia, Univerist\`{a} degli Studi di Perugia, Italy}
\affiliation{INFN, Sezione di Perugia, Via A. Pascoli, I-06123, Perugia,  
Italy}
\author{\textsc{J. T. Gaglione}}
\affiliation{Department of Physics and Astronomy, Vanderbilt University, Nashville, TN, 37235, USA}
\author{\textsc{A.~Gurrola}}
\affiliation{Department of Physics and Astronomy, Vanderbilt University, Nashville, TN, 37235, USA}
\author{\textsc{O.~Panella}}
\affiliation{INFN, Sezione di Perugia, Via A. Pascoli, I-06123, Perugia,  
Italy}
\author{\textsc{M. Presilla}}
\affiliation{Institute for Experimental Particle Physics (ETP), Karlsruhe Institute of Technology (KIT), Wolfgang-Gaede-Straße 1, 76131 Karlsruhe, Germany}

\author{\textsc{F. Romeo}}
\affiliation{Department of Physics and Astronomy, Vanderbilt University, Nashville, TN, 37235, USA}
\author{\textsc{H. Sun}}
\affiliation{Institute of Theoretical Physics, School of Physics, Dalian University of Technology, No.2 Linggong Road, Dalian, Liaoning, 116024, P.R.China}

\author{\textsc{S.-S. Xue}}
\affiliation{ICRANet, Piazzale della Repubblica, 10-65122, Pescara, Italy}
\affiliation{Physics Department, Sapienza University of Rome, Piazzale Aldo Moro 5, 00185 Roma, Italy}
\affiliation{INFN, Sezione di Perugia, Via A. Pascoli, I-06123, Perugia,  
Italy}
\affiliation{ICTP-AP, University of Chinese Academy of Sciences, Beijing, China}

\begin{abstract}
We present a detailed study concerning a new physics scenario involving four fermion operators of the Nambu-Jona-Lasinio type characterized by a strong-coupling ultraviolet fixed point where composite particles are formed as bound states of elementary fermions at the scale $\Lambda ={\cal O}(\text{TeV})$. After implementing the model in the Universal FeynRules Output format, we focus on the phenomenology of the scalar leptoquarks at the LHC and the High-Luminosity option. Leptoquark particles have undergone extensive scrutiny in the literature and experimental searches, primarily relying on pair production and, more recently, incorporating single, Drell-Yan t-channel, and lepton-induced processes. This study marks, for the first time, the examination of these production modes at varying jet multiplicities. 
Novel mechanisms emerge, enhancing the total production cross section. 
A global strategy is devised to capture all final state particles produced in association with leptoquarks or originating from their decay, which we termed ``exclusive'', in an analogy to the nomenclature used in nuclear reactions. The assessment of the significance in current and future LHC runs, focusing on the case of a leptoquark coupling to a muon - \textit{c} quark pair, reveals greater sensitivity compared to ongoing searches. Given this heightened discovery potential, we advocate the incorporation of exclusive leptoquark searches in future investigations at the LHC.
\end{abstract}

\maketitle
%

\setcounter{footnote}{0}

\newpage
\tableofcontents
\newpage

\section{Introduction}
Thanks to its remarkably strong quantitative agreement with experimental results over the past decades, the Standard Model (SM) has been solidified as a cornerstone of elementary particle. Nonetheless, it is regarded as an effective theory that falls short in encompassing several foundational aspects of both theoretical and observed particle physics. Numerous extensions have been put forward, suggesting specific tests to challenge the SM at the LHC. Regrettably, all of these searches have thus far yielded a negative outcome in the quest for detecting new physics and have placed signiﬁcant constraints on many beyond-the-SM (BSM) models in mass scales up to a few TeV~\cite{ATLAS:2019fgd, ATLAS:2021jyv, ATLAS:2019isd,CMS:2022zoc,CMS:2021ctt,CMS:2019gwf,CMS:2020wzx,CMS:2022rqc}.

Therefore, the question arises as to whether ongoing searches have exhaustively explored all avenues, or if unprecedented aspects can yet be harnessed to devise better methods that would probe yet uncharted territory. In this paper, we present an in-depth phenomenological study of a novel composite scenario based on Nambu-Jona-Lasinio (NJL) type four-fermion operators recently proposed in~\cite{Xue:2016dpl, Xue2017,Xue2023}. The model is characterized by a strong coupling ultraviolet (UV) fixed point where composite particles arise as bound states of SM elementary fermions.  The counterpart of a strong-coupled electron system is studied \cite{Kleinert:2017nld} in condensed-matter physics.
The effective four-fermion interactions considered in this new model are motivated by theoretical inconsistencies~\cite{Nielsen:1980rz,Nielsen:1981hk,Nielsen:1991si} between the SM bilinear Lagrangian of chiral gauged fermions and the UV regularization of some unknown dynamics, such as quantum gravity. These inconsistencies imply, at high energies, quadrilinear effective operators (four-fermion interactions) of the NJL or Einstein-Cartan type~\cite{Xue:2008qs, Xue:2009zza, Xue:2009wm}. 
A first phenomenological study related to the production at the LHC of composite fermions in the gauge symmetric phase of this NJL model was conducted in~\cite{Leonardi:2018jzn} where a recent CMS analysis~\cite{CMS:2017sqy} of the dilepton plus dijet final state was recast to derive constraints on the parameter space of the NJL model. More recent studies of the XENON1T experiment~\cite{Shakeri:2020wvk}, neutrino-less double beta decay~\cite{Pacioselli_2023},
and $W$-boson mass tension~\cite{Xue:2022mde} addressed the broken-gauge phase of the model. In this work we specifically address the phenomenology at the LHC of the scalar leptoquarks (LQs) arising in the gauge symmetric phase of the model, thus delving into one of the most extensively studied subjects in the BSM landscape. While reviewing current investigations, production modes, and search strategies, we investigate the potential for new mechanisms and approaches at the LHC.

LQs are hypothetical particles that carry both lepton number $L$ and baryon number $B$, have a fractional electric charge, and can be either scalar (spin 0) or vector (spin 1) particles. They offer a natural explanation for the symmetry between 
the quark and lepton families, attempting to pave the way toward matter unification. They inherently appear in many well-motivated extensions of the SM that endeavor to unify the fundamental interactions, such as grand unified theories~\cite{Pati:1973uk, PatiSalam, GeorgiGlashow, Fritzsch:1974nn, GERSHTEIN1999159, Dorsner:2005fq}, technicolor models~\cite{Dimopoulos:1979es,Dimopoulos:1979sp,Technicolor, Lane:1991qh}, and compositeness scenarios~\cite{LightLeptoquarks,Gripaios:2009dq}, as well as $R$-parity violating supersymmetry~\cite{Farrar:1978xj,Ramond:1971gb,Golfand:1971iw,Neveu:1971rx,Volkov:1972jx,Wess:1973kz,Wess:1974tw,Fayet:1974pd,Nilles:1983ge,Barbier:2004ez}, and as mediators of dark matter-SM interactions~\cite{Baker:2015qna}. More recently, LQs have gained enhanced interest, as they provide a suitable explanation for a series of anomalies observed in several precision measurements: the $B\rightarrow D^*$ anomaly reported by BaBar~\cite{BaBar:2012obs, BaBar:2013mob}, Belle~\cite{Belle:2015qfa}, and LHCb~\cite{LHCb:2015gmp} that could be explained by the mediation of an intermediate LQ scalar~\cite{Mandal2019, Dumont:2016xpj, Chen:2017hir, Barbieri:2016las, Tanaka:2012nw, Iguro:2023prq, Crivellin:2017zlb, Buttazzo:2017ixm, DiLuzio:2017vat, Bordone:2017bld, Greljo:2018tuh, DiLuzio:2018zxy, Fuentes-Martin:2020bnh}; the muon anomalous magnetic moment $(g-2)_\mu$ measurement from the E989 and E821 collaborations~\cite{Muong-2:2021ojo, Muong-2:2006rrc} which could be resolved assuming the existence of LQs~\cite{Bennett2006, Muong-2:2021ojo, Bauer:2015knc, Greljo:2021xmg, Du:2021zkq, Parashar:2022wrd, Chen:2022hle, Chowdhury:2022dps, He:2022zjz, Cheung:2022zsb, He:2021yck, Bigaran:2021kmn, Wang:2021uqz, Zhang:2021dgl, FileviezPerez:2021lkq, Ban:2021tos,ColuccioLeskow:2016dox,Crivellin:2020tsz}; the charged proton radius obtained from muon Lamb shift \cite{Pohl2010}, and atomic parity violation \cite{Wood1997, Dorsner2014}, where low-energy corrections via virtual LQ scalars and internal loop diagrams can contribute. In a search for a LQ coupling to a $\tau$ lepton and a $b$ quark, the CMS collaboration has reported an excess with a significance of 2.8 standard deviations above the SM expectation using Run 2 data~\cite{CMS:2023qdw}, though consistency with the SM prediction is observed in a similar search from the ATLAS collaboration~\cite{ATLAS:2023vxj}. 

\begin{figure}[!h]
\centering
\subfloat[]{\includegraphics[scale=0.265]{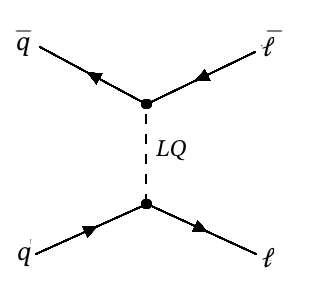}}
\subfloat[]{\includegraphics[scale=0.265]{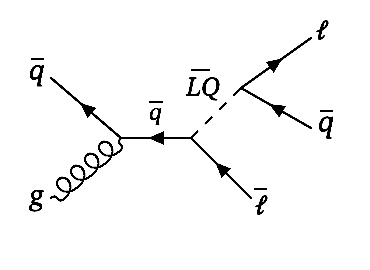}}
\subfloat[]{\includegraphics[scale=0.265]{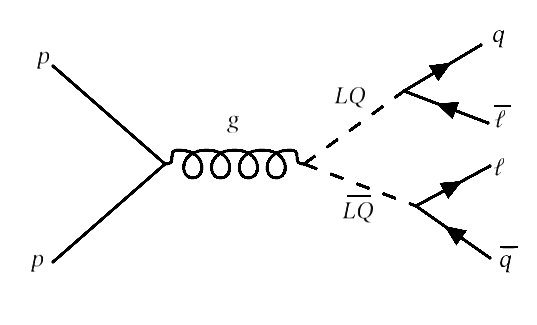}}
\subfloat[]{\includegraphics[scale=0.265]{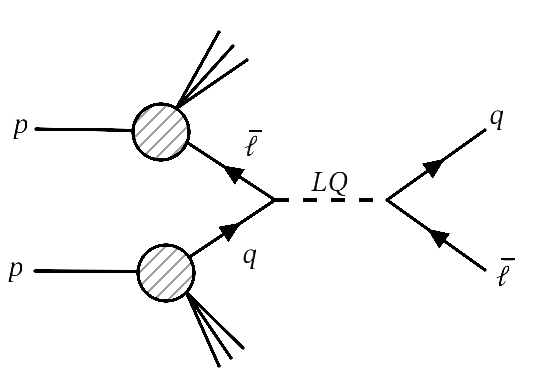}}
\caption{Examples of Feynman diagrams for the Drell-Yan t-chnnel (a), single (b), pair (c), and lepton-induced (d) production modes considered at the LHC.}
\label{LQ_established_processes}
\end{figure}

Both the ATLAS and the CMS collaborations have a broad LQ search program that covers various particle hypotheses and different lepton-quark-LQ vertex couplings $\lambda$. The main production modes considered are illustrated in Fig.\ref{LQ_established_processes} (a - d) and have been extensively discussed in literature~\cite{Diaz:2017lit, Blumlein:1996qp, Dorsner:2022ibm, Dorsner:2021chv, Borschensky:2022xsa, Ohnemus:1994xf, Mandal:2015vfa, Raj:2016aky, Haisch:2022lkt, Haisch:2022afh, Bansal:2018eha, Schmaltz:2018nls, Greljo:2020tgv, Buonocore:2020erb, Buonocore:2022msy, Korajac:2023xtv, Florez:2023jdb, Crivellin:2021ejk}. The majority of the analyses rely on the LQ pair production mode in Fig.\ref{LQ_established_processes} (c)~\cite{ATLAS:2023prb, ATLAS:2023kek, ATLAS:2023uox, ATLAS:2022wcu, ATLAS:2021jyv, ATLAS:2021yij, ATLAS:2021oiz, ATLAS:2020xov, ATLAS:2020dsk, ATLAS:2020dsf, ATLAS:2019qpq, ATLAS:2019ebv, ATLAS:2016wab, ATLAS:2015hsi, ATLAS:2013oea, ATLAS:2012aq, ATLAS:2011zhi, ATLAS:2011atv, CMS:2018ncu, CMS:2022nty, CMS:2018lab, CMS:2018oaj, CMS:2022nty, CMS:2018iye, CMS:2022nty, CMS:2019ybf, CMS:2018yiq, CMS:2017xcw, CMS:2016fxb, CMS:2015gua, CMS:2015nep, CMS:2014wpz, CMS:2012cyn, CMS:2012iln, CMS:2011zfm, CMS:2010ssz, CMS:2010chx}, which does not depend on $\lambda$. Some searches are designed to intercept the single LQ production process in Fig.\ref{LQ_established_processes} (b)~\cite{ATLAS:2023vxj, ATLAS:2021tar, CMS:2015xzc, CMS:2018txo} or the single and pair productions together~\cite{CMS:2020wzx, CMS:2021far, ATLAS:2023vxj}, as
the former brings enhanced sensitivity due to a cross section that varies as a function of ${\lambda}^2$. A recent investigation~\cite{CMS-PAS-EXO-19-016} added further Drell-Yan dilepton t-channel contributions as displayed in Fig.\ref{LQ_established_processes} (a) and ~\cite{CMS:2022ncp} was designed specifically for this process, whose cross section depends on ${\lambda}^4$. An additional recent search relies on lepton-induced processes~\cite{CMS:2023bdh}, as that in Fig.\ref{LQ_established_processes} (d). \\
Upon reviewing the existing literature and LHC searches, a noteworthy observation emerges: the generation of LQs via t-channel, single, and pair production processes is solely regarded in terms of their primary production modes. No studies have been carried out to explore the implications of these production mechanisms at various jet multiplicities, assuming that a jet is originating from either a gluon or a quark. Nevertheless, previous research~\cite{CMS:2023byi, ATLAS:2021mla, ATLAS:2019odt, CMS:2019mij, Abdullah:2018ets, Abdullah:2017oqj, Barbosa:2022mmw, Abdullah:2019dpu} has demonstrated that such an approach could amplify the discovery potential by incorporating additional signal sources while effectively reducing the SM background. In this work, we pursue this effort and investigate how LQs would manifest in association with jets. Our investigation uncovers new production mechanisms and topological aspects not yet considered in ongoing searches. Furthermore, we devise a comprehensive strategy to look for LQs, taking into account all signal contributions. For the case of the lepton-induced process, we undertake a study to separate elastic and inelastic contributions. Our findings reveal that the sensitivity achieved surpasses that of all ongoing searches. 
This new approach developed here for the specific case of LQs can be adapted and applied in searches for any type of particle to enhance further the expected sensitivity. \\
In Section~\ref{leptoquark_model}, we introduce the reference LQ model utilized in this analysis, while in Section~\ref{signal_sample_generation} we delve into its implementation and the generation of the signal samples. Section~\ref{exclusive_lq_production} describes the LQ production modes in conjunction with jets and touches upon the possibility of distinguishing elastic and inelastic lepton-induced processes. In Section~\ref{search_strategy} and Section~\ref{expected_sensitivity}, we present a global strategy for the detection of LQs based on our findings, as well as the anticipated sensitivity results at both the LHC and the High-Luminosity (HL)-LHC. Finally, in Section~\ref{summary_remarks}, we summarize the work with some closing remarks.

\section{Nambu-Jona-Lasinio composite leptoquark model}\label{leptoquark_model}

The No-Go theorem \cite{Nielsen:1980rz,Nielsen:1981hk,Nielsen:1991si}
shows that, as a renormalizable quantum field theory at infrared (IR) low energies, the SM {\it bilinear} fermion Lagrangian at UV cutoff $\Lambda_{\rm cut}$ suffers theoretical inconsistencies with SM chiral gauge symmetries and fermion spectra at low energies. This implies that new physics at the UV scale $\Lambda_{\rm cut}$ effectively presents {\it quadrilinear} operators $G_{\rm cut}\bar\psi^{f}_{_L}\psi^{f}_{_R}\bar\psi^{f}_{_R} \psi^{f}_{_L}$  ($G_{\rm cut}\propto \Lambda^{-2}_{\rm cut}$) \cite{Xue:2016dpl, Xue2017} of Nambu-Jona-Lasinio (NJL) or Einstein-Cartan type.
They are SM gauge symmetric interactions of SM left- and right-handed fermions ($\psi^f_L,\psi^f_R$) in the family ``$f$''.
The strong coupling $g_{\rm cut }= G_{\rm cut}\Lambda^2_{\rm cut}>1$ dynamics leads to massive composite bosons $\Pi^{f}\sim (\bar\psi^{f}_{_R} \psi^{f}_{_L})$ and fermions $F^{f}_R\sim \psi^{f}_{_R} (\bar\psi^{f}_{_R} \psi^{f}_{_L})\sim \psi^{f}_{_R}\Pi^{f}$ \cite{Xue1997, Xue2000, Xue2000a}. They carry SM charges and couple to massless SM gauge bosons in a gauge-invariant manner. 
The effective SM gauge symmetric and renormalizable theory realizes in the scaling domain of the UV fixed point $g^*_{\rm cut}$ at the composite scale $\Lambda\ll \Lambda_{\rm cut}$ \cite{Xue2014, Xue2017}.  
As the energy scale decreases below $\Lambda$, the effective theory undergoes \cite{Xue2023} a phase transition of composite particle decays into constituents and spontaneous SM symmetries breaking via the top-quark channel $(\bar t_Lt_R\bar t_R t_L)$, leading to the SM at the electroweak scale $v\approx 246$ GeV \cite{Bardeen1990}. The scale $\Lambda = {\cal O}(\text{TeV})$ is estimated by using the top quark and Higgs boson masses and extrapolating the renormalization-group solution to the high-energy regime \cite{Xue2013}.

\begin{table*}
\begin{tabular}{cccccc}
Bosons $\Pi_a^Q$ &  Charge $Q_i=Y+t^i_{3L}$, & $SU_L(2)$ 
3-isospin $t^i_{3L}$,& $U_Y(1)$ charge $Y$, & $SU_c(3)$ color $a$,& LQ \cr
\hline\hline
$\Pi^{+5/3}_a \propto \bar e_{R}u_{La} $& $+5/3$& $+1/2$ &$+7/6$ & 3 & $R_2^{+5/3}$ \cr
$\Pi^{-1/3}_a \propto \bar \nu^e_{R}d_{La} $  &$-1/3$& $-1/2$& $+1/6$& 3& $\tilde R_2^{-1/3}$\cr
$\Pi^{+2/3}_{u_a} \propto \bar \nu^e_{R}u_{La}$ &$ +2/3$ &$+1/2$ &$+1/6$& 3& $\tilde R_2^{+2/3}$\cr
$\Pi^{+2/3}_{d_a} \propto \bar e_{R}d_{La}$ &$+2/3$ & $-1/2$ & $+7/6$& 3& $R_2^{+2/3}$\\
\hline
$\Pi^{-5/3}_a \propto \bar u_{Ra}e_{L}$  & $-5/3$& $-1/2$ &$-7/6$& 3& $R_2^{-5/3}$\cr
$\Pi^{+1/3}_a \propto \bar d_{Ra} 
\nu^e_{L}$  &$+1/3$& $+1/2$& $-1/6$& 3& $\tilde R_2^{+1/3}$\cr
$\Pi^{-2/3}_{u_a} \propto \bar u_{Ra}\nu^e_{L}$ &$- 2/3$ &$+1/2$ &$-7/6$& 3& $R_2^{-2/3}$\cr
$\Pi^{-2/3}_{d_a}\propto \bar d_{Ra}e_{L}$  & $-2/3$ &$-1/2$ &$-1/6$& 3& $\tilde R_2^{-2/3}$\\
\hline\hline
\end{tabular}
\caption{Composite bosons, their constituents, standard model charges, and corresponding leptoquark according to the nomenclature used in~\cite{Dorsner2016, Dorsner:2018ynv} for the first generation of standard model fermions.} 
\label{qnuP0ql}
\end{table*}

Composite bosons $\Pi^f$ and fermions $F^f$ spectra at the $\Lambda$ scale offer a vast range of new searches and tests of the SM, as shown in~\cite{Leonardi:2018jzn}. In this article, we study the phenomenological analysis of composite boson states $\Phi\sim (\bar\ell_{R}q_{L})$ with fractional electric charge, or LQ, as listed in Table \ref{qnuP0ql}.
Their SM gauge charges are the sum of the SM gauge charges of their fermionic constituents. The gauge-invariant kinetic Lagrangian is
\begin{eqnarray}
(D_\mu \Phi)^\dagger (D_\mu \Phi) + M^2_\Pi \Phi^\dagger \Phi,
\label{kinetic}
\end{eqnarray}
with covariant derivative,
\begin{eqnarray}
 D_{\mu }=\partial_ {\mu  }+\text{ig}_1 Y B_{\mu }+\frac{1}{2} \text{ig}_2 \sigma _iW^i{}_{\mu }+\text{ig}_3 T^aG^a{}_{\mu }
 \label{covd}
 \end{eqnarray}
where $Y$ is the LQ hypercharge, $\sigma _i$ are the standard Pauli matrices and $T^a$ are the Gell-Mann matrices. Moreover, $g_1$, $g_2$, and $g_3$ are the SM gauge couplings for the respective group.  
In Eq.~\ref{kinetic} $SU_c(3)$ triplet LQ bosons form $SU_L(2)$ doublets. Identification of doublets in terms of hypercharge is given as $\Pi^a_{1/6}$= $\begin{pmatrix}
 \Pi _{u_{a}}^{2/3} \\
 \Pi _{a}^{-1/3} 
\end{pmatrix}$ and $\Pi^a_{7/6}$= $\begin{pmatrix}
 \Pi _{a}^{5/3} \\
 \Pi _{d_{a}}^{2/3} 
\end{pmatrix}$. 
The effective coupling between a LQ composite boson 
and its two constituents can be written as effective contact interactions,
\begin{equation}
 {\mathcal L}_{CI} = g_{_{\Pi_{5/3}}} (\bar e_{R}u_{La})\Pi^{-5/3}_a  +  g_{_{\Pi_{1/3}}}(\bar\nu^e_{R}d_{La})\Pi^{1/3}_a + 
 g_{_{\Pi_{-2/3}}} (\bar \nu^e_{R}u_{La}) \Pi^{-2/3}_{u_a} +g_{_{\Pi_{-2/3}}}(\bar e_{R} d_{La})\Pi^{-2/3}_{d_a} + {\rm h.c.}  \label{iqlBdql1}
\end{equation}
where the LQ effective Yukawa coupling $g_{_{\Pi_{i}}}=(F_{\Pi_{i}}/\Lambda)^2\sim {\mathcal O}(1)$. The $g_{_{\Pi_{i}}}$ parameter defines the decay width (as calculated in ~\cite{Leonardi:2018jzn} for composite bosons with integer charge) via the formula $\Gamma(q\ell) = (g_{_{\Pi_{i}}}^2 m_{\Pi})/16\pi$ that coincides with the general formula used for scalar LQ (see~\cite{Dorsner:2016wpm}). For the sake of simplicity and to align to the notation commonly used in experiments (see e.g.~\cite{CMS:2020wzx,CMS:2023qdw}) we will use $\lambda$ to stand for $g_{_{\Pi_{i}}}$ in the sections below. 
The effective interacting Lagrangian 
(\ref{kinetic}-\ref{iqlBdql1}) at the scale $\Lambda$ is dimension-4 renormalizable, namely independent from the cutoff $\Lambda_{\rm cut}$. 
We tabulate the four quark-lepton composite bosons $\Pi^{5/3}_a$, $\Pi^{-1/3}_a$, $\Pi^{2/3}_{u_a}$ and 
$\Pi^{2/3}_{d_a}$, identified by  
constituent fermions and SM gauge charges. Their conjugated fields are $(\Pi^{5/3}_a)^\dagger=\Pi^{-5/3}_a$, $(\Pi^{-1/3}_a)^\dagger=\Pi^{1/3}_a$ and $(\Pi^{2/3}_{u_a})^\dagger=\Pi^{-2/3}_{d_a}$, and $(\Pi^{2/3}_{d_a})^\dagger=\Pi^{-2/3}_{u_a}$.
The renormalized LQ bosons $\Pi^Q_a=Z_\Pi^{-1/2}\bar \ell_{R}q_{La}$ with the 
form-factor $Z_\Pi^{-1/2}=(\sqrt{4\pi}/F_\Pi)^2$ have the same dimension $[energy]$ of elementary boson fields. Their masses $M_\Pi\propto\Lambda$ and decay constants $F_\Pi\propto\Lambda$, due to the composite dynamics at the scale $\Lambda$.
These LQ states have baryon number $B=1/3$ and lepton number $L=-1$, corresponding to $R_2$ and $\tilde R_2$ states in the LQ nomenclature \cite{Dorsner2016, Dorsner:2018ynv}, which do not suffer interference with SM processes as pointed out in~\cite{Bansal:2018eha}.
The spectra in Table (\ref{qnuP0ql}), interactions (\ref{iqlBdql1}), can be generalized by
$\nu^e,e,u,d \rightarrow \nu^\mu,\mu,c,s$ for the second generation, and $\nu^e,e,u,d \rightarrow \nu^\tau,\tau,t,b$ for the third generation.  

Although coupling $G_{\rm cut}$ and the composite scale $\Lambda$ are unique, the coupling $g_{_{\Pi_i}}$ and mass $M_{\Pi_i}$ have different values but are similar and related across generations.
The main reason is that, in the four-fermion interactions, the fermion fields $\psi^f_{L,R}$ are SM gauge eigenstates. While in the ground state, they mix in different charged sectors and generations by unitary transformations $U^{\nu,\ell, u,d}_{L,R}$ \cite{Xue:2016dpl}. 
Therefore, LQ states in different generations mix, provided they have the same SM charges of Table \ref{qnuP0ql}. For example,  $\Pi^{5/3}_{a}(\bar e_{R}u_{La})$ exhibits mixing amongst the LQ states $(\bar \mu_{R}c_{La})$ and $(\bar \tau_{R}t_{La})$. 
In addition to the Yukawa coupling (\ref{iqlBdql1}), it can also couple to the lepton and quark in the second and third generations via flavor-mixed Yukawa couplings,
\begin{eqnarray}
g_{_{\Pi_{5/3}}} (U^\dagger_R U_L)_{1,2} (\bar \mu_{R} c_{La} )\Pi^{-5/3}_a 
+g_{_{\Pi_{5/3}}} (U^\dagger_R U_L)_{1,3} (\bar \tau_{R} t_{La} )\Pi^{-5/3}_a,
\label{ciqlBqlmix}
\end{eqnarray}
where the flavor-suppressed factors $(U^\dagger_R U_L)_{1,3}<(U^\dagger_R U_L)_{1,2} < 1$ 
stand for the CKM-like mixing matrix elements between the first and other generations. The same discussions apply to other LQ states in Table \ref{qnuP0ql}. Other flavor-mixed Yukawa couplings are possible, as long as they preserve SM gauge symmetries. 
For these reasons, LQ boson masses and Yukawa couplings should be different in generations and related by an unknown mixing matrix $(U^{\nu,\ell, u,d}_{L, R})^\dagger U^{\nu,\ell, u,d}_{L, R}$, in analogy to the CKM matrix in the quark sector and PMNS matrix in the lepton sector. 
We treat different composite boson masses $M_{\Pi_i}$ 
and Yukawa couplings $g_{_{\Pi_i}}$ as theoretical parameters. The constraint $g^2_{_{\Pi_i}}<8\pi/\sqrt{3}$ is inferred by studying the unitary problem \cite{DiLuzio:2017chi}.

These LQ bosons (Table \ref{qnuP0ql}) and interactions (\ref{covd}-\ref{iqlBdql1}) are the counterparts of color singlets $\Pi^{0,\pm}$ in Table V and interactions (26-28) of Ref.~\cite{Leonardi:2018jzn} by exchanging $e_{L,R}\leftrightarrow d^a_{L,R}$ and $\nu^e_{L,R}\leftrightarrow u^a_{L,R}$.  
The parameter spaces of Yukawa couplings $g_{_{\Pi_i}}$ and masses $M_{\Pi_i}$ for investigating $\Pi^{0,\pm}$ are the same as those of LQ bosons up to the aforementioned mixing matrix elements of the order of unity. The differences are only the initial states' parton density functions (PDFs) used and the final states detected. For color singlets $\Pi^{0,\pm}$, there are decay (production) channels to (from) SM gauge bosons $W,Z$ and $\gamma$, while color LQ states can be produced by gluon splitting or via a lepton-quark vertex. Analogously to the studies of composite fermions \cite{Leonardi:2018jzn}, we will discuss composite LQ fermions in a separate article, e.g. $F_{R}\sim \ell_{R}\Phi$ or $q_{R}\Phi$, interacting with SM gauge bosons and decaying to $\ell_{R} (q_{R})+\Phi\rightarrow \ell_{R} (q_{R}) + \bar\ell_{R}+ q_L$. 
{\textcolor{black}We also mention the possibility of nontrivial contributions to 
the $B\rightarrow D^*$ anomaly $R_{D^{(*)}}$ of the charged current channel 
$b \rightarrow c \tau \bar\nu$ based on the composite boson spectra (Table \ref{qnuP0ql}). Having the same SM charges, the state $\Pi^{-2/3}_{u_a}(\bar u_{Ra}\nu^e_{L})$ mixes with the states $(\bar c_{Ra}\nu^\mu_{L})$, and the state $\Pi^{2/3}_{d_a}(\bar e_{R} d_{La})$ mixes with the states $(\bar\tau_{R} b_{La})$. Therefore, we have the transition amplitude $\langle [\Pi^{2/3}_{u_a}(0)]^\dagger\Pi^{2/3}_{d_a}(x)\rangle\rightarrow ({\mathcal C}/\Lambda^2) \langle (\bar c_{Ra}\nu^\mu_{L})(\bar\tau_{R} b_{La})\rangle$, yielding to the probability of the flavour-changing process $b \rightarrow c \tau \bar\nu$, where the ${\mathcal C}$ relates to the mixing amongst quark and lepton flavours. This is similar to the discussion of the LQ scalar $R^{+2/3}_2$ ($\Pi^{+2/3}_{d_a}$) contribution to $R_{D^{(*)}}$ \cite{Angelescu_2021}, see also \cite{Buttazzo_2017}. We leave for a future effort the detailed analysis in this context.} 

\section{Signal implementation and samples generation}
\label{signal_sample_generation}
To generate signal and background events, we utilize MadGraph5 aMC@NLO\cite{Alwall:2011uj}, incorporating the gauge (Eq.~\ref{kinetic}) and the contact (Eq.~\ref{iqlBdql1}) interaction terms described in the previous section via the Feynrules package \cite{Alloul:2013bka} as a Universal FeynRules Output (UFO) module \cite{Degrande:2011ua}. Subsequently, we generate Monte Carlo (MC) samples for both the signal and SM background processes via MadGraph5 aMC@NLO, then use 
Pythia 8.2 \cite{Sjostrand:2014zea} for including initial and final parton showering and fragmentation. The Madgraph implementation includes the \textit{b} quark in our proton definition, and the interactions have been implemented up to the 3rd fermion generation with the corresponding couplings.
\\ 
In this article, we consider LQs with a charge of $5/3$ that couple to electron-\textit{u} quark ($\lambda_{e u}$) and muon-\textit{c} quark ($\lambda_{\mu c}$) pairs for the signal generation, {\color{black} along with an LQ with a charge of $2/3$ that couples to tau-\textit{b} quark} ($\lambda_{\tau b}$). The signal process is generated with two types of beam hypotheses: proton-proton ($pp$) and photon-proton ($\gamma p$). For the $pp$ LQ production, the NNPDF31-lo-as-0118~\cite{NNPDF:2017mvq} PDF is used. For the $\gamma p$ collisions, we choose the MRST2004qed-proton PDF~\cite{Martin:2004dh}. This decision is motivated by our aim to distinguish between the elastic and inelastic contributions of the photon. The NNPDF set previously mentioned incorporates both the elastic and inelastic contributions (as does the LUX PDF~\cite{Bertone:2017bme}), rendering it impractical to separate the two components. The MRST2004qed PDF allows for the separation of the elastic and inelastic contributions with the configurations available in Madgraph. 
In our studies, we simulate the processes $pp \rightarrow \ell^+ \ell^-$ plus 0, 1, 2, or at least 3 partons to incorporate the standard production modes in Fig.~\ref{LQ_established_processes} (a - c) and the new production modes that are presented in the subsequent section. For the photon-induced processes, we consider $\gamma p \rightarrow \ell^+ \ell^-$ plus 1 parton to focus our study on elastic versus inelastic contribution only. \\
The center-of-mass energy is set to 13 TeV. The values of $\lambda$ vary in the range [0.5-2.5] in steps of 0.5. For the LQ masses, we consider values in the ranges (in TeV) [1.00-2.00] in steps of 0.25, [2.00, 10.00] in steps of 0.50, plus 20.00 and 30.00 TeV. 
Extra jets in the production of the LQ may originate from either gluons or quarks, as specified below. The former corresponds to initial or final state radiation emitted in one of the well-searched, tree-level diagrams, while the latter may give rise to novel mechanisms to generate LQs. We note that both scenarios can contribute to the overall signal, necessitating the avoidance of double counting of events among different processes. To achieve this, in the production of the signal samples, we employ the matrix-element parton shower matching technique known as MLM matching~\cite{Mangano:2006rw}. We set a matching scale of 30 GeV and ensure appropriate overlap removal in the differential jet rates of our samples. 
We have verified that the cross sections obtained using our implemented model are consistent within 1\% with those yielded by the models LQ-UFO~\cite{Lqufo}, commonly used in analyses at LHC~\cite{CMS:2020wzx}, 
using the same generator configurations and relying on the resonant pair production mode (Fig.~\ref{LQ_established_processes} (c)) that depends on $\alpha_s$ only.

\section{Exclusive LQ production}
\label{exclusive_lq_production}
In this section, we explain LQ production in detail, emphasizing the mechanisms where LQs appear with additional jets compared to the well-established modes depicted in Fig.~\ref{LQ_established_processes}. We refer to this as ``exclusive'' LQ production, drawing an analogy to the nomenclature utilized in the study of inelastic scattering, where every reaction product is interested. Our examination of the numerous Feynman diagrams reveals three distinct cases of interest: 
(1) the splitting of a gluon into two quarks, one of which participates in the LQ generation, the other appearing as an additional jet;
(2) the emission of a gluon from a parton that contributes to the LQ formation through separation into two quarks, with the original radiating parton and one of the gluon's quarks ending as additional jets; 
(3) the exchange of a lepton that may lead to up to two additional jets. 
We remark that, although we are highlighting these possibilities in the context of the LQ particle, 
they apply to any SM and BSM particle that permits the relevant couplings to quarks and leptons. It is also possible to have the emission of a gluon, arising from initial or final state radiation, that subsequently manifests as an additional jet (or two jets if it divides into quarks) and does not couple to the LQ. Although this latter occurrence also augments the jet multiplicity compared to the reference modes at the Born level, this scenario does not introduce new production processes, as it is anticipated within the hadronic environment of the LHC for any mechanism. 
The third instance above has been previously discussed in relation to pair production~\cite{Dorsner:2022ibm, Dorsner:2021chv} only, although it may occur also in t-channel and single production, and it introduces two additional new physics vertices that enhance the cross section rapidly with respect to $\lambda$. 
The generation of an LQ initiated via a lepton is a unique production mode, and we treat it in the next sub-section.

\begin{figure}[!h]
\centering
\subfloat[]{\includegraphics[scale=0.2]{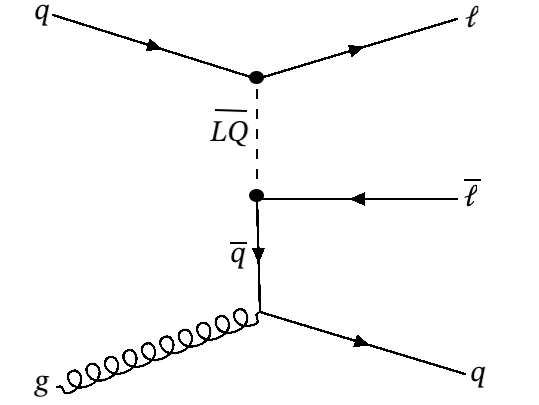}}
\subfloat[]{\includegraphics[scale=0.17]{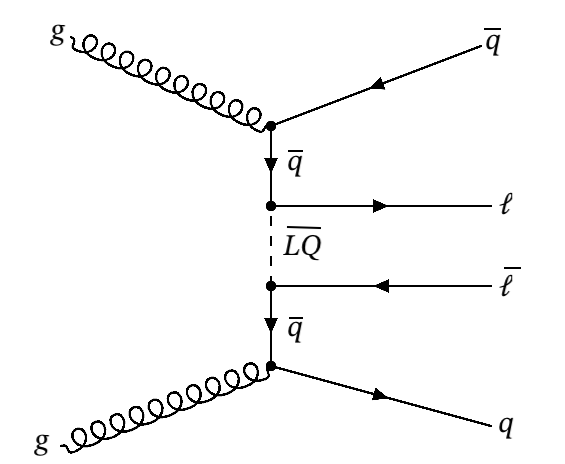}}
\subfloat[]{\includegraphics[scale=0.2]{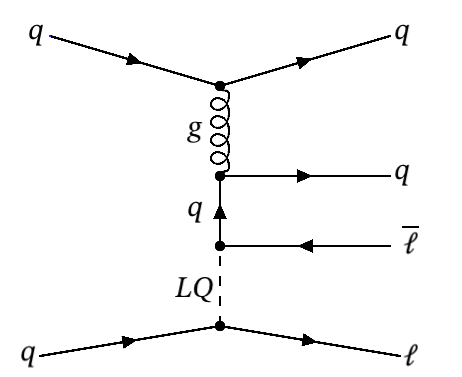}}
\subfloat[]{\includegraphics[scale=0.2]{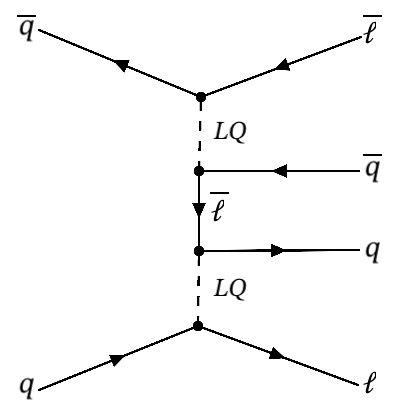}}\\\vspace{-0.5cm}
\subfloat[]{\includegraphics[scale=0.17]{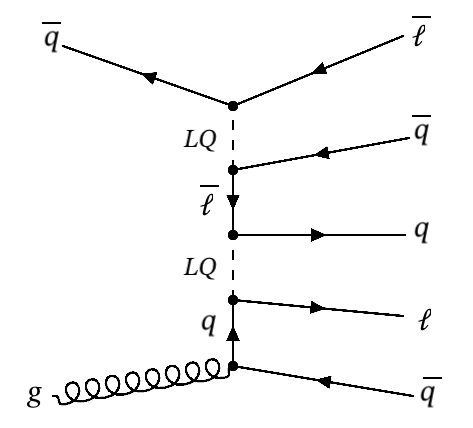}}
\subfloat[]{\includegraphics[scale=0.15]{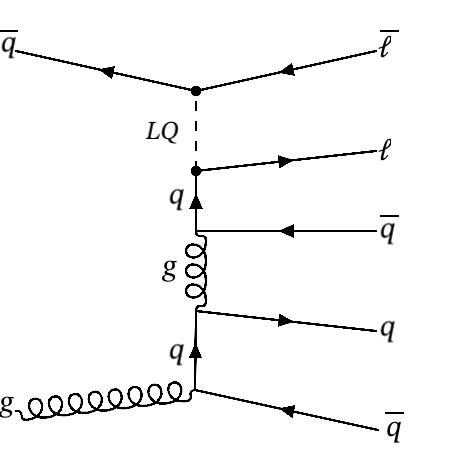}}
\subfloat[]{\includegraphics[scale=0.2]{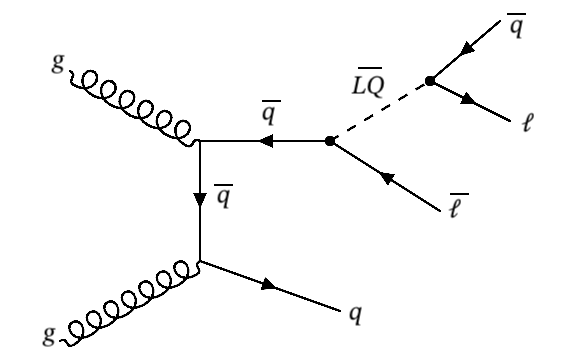}}
\subfloat[]{\includegraphics[scale=0.2]{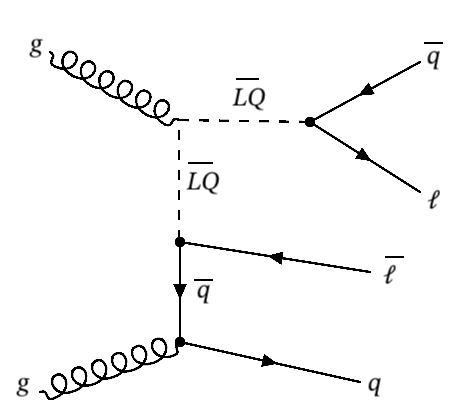}}\\\vspace{-0.5cm}
\subfloat[]{\includegraphics[scale=0.2]{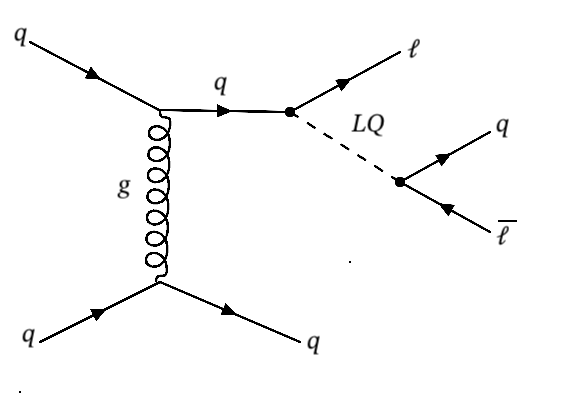}}
\subfloat[]{\includegraphics[scale=0.2]{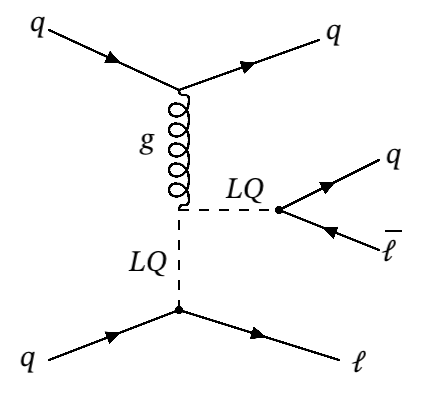}}
\subfloat[]{\includegraphics[scale=0.2]{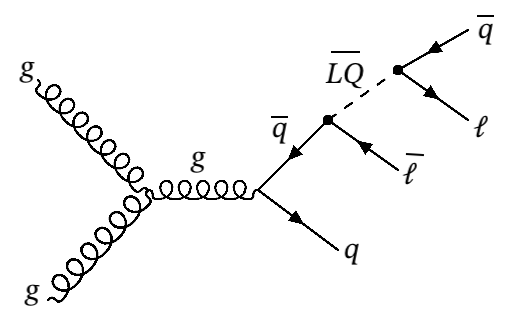}}
\subfloat[]{\includegraphics[scale=0.2]{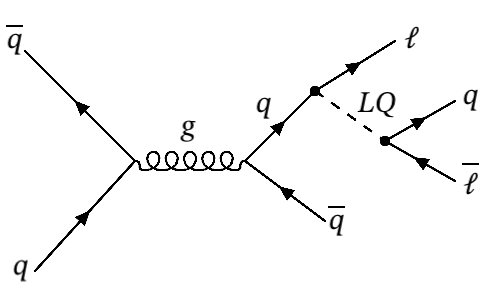}}\\\vspace{-0.5cm}
\subfloat[]{\includegraphics[scale=0.17]{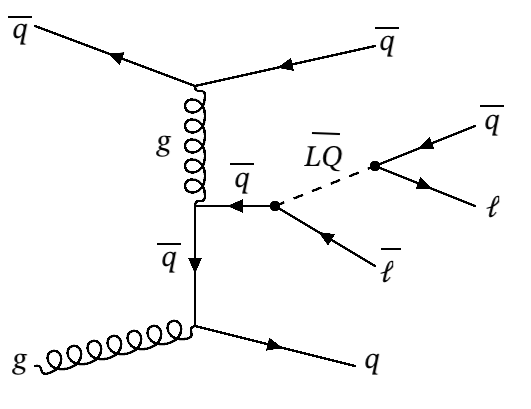}}
\subfloat[]{\includegraphics[scale=0.17]{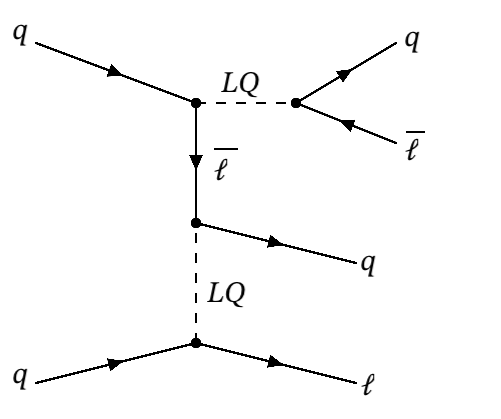}}
\subfloat[]{\includegraphics[scale=0.17]{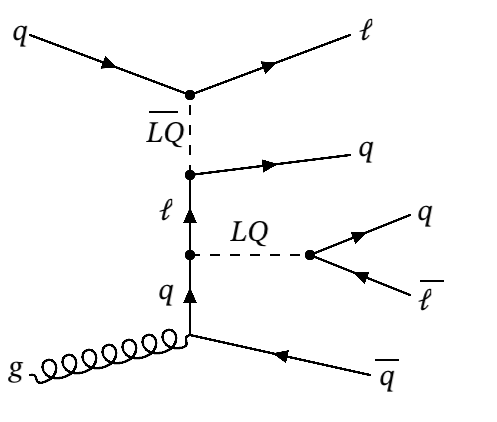}}
\subfloat[]{\includegraphics[scale=0.2]{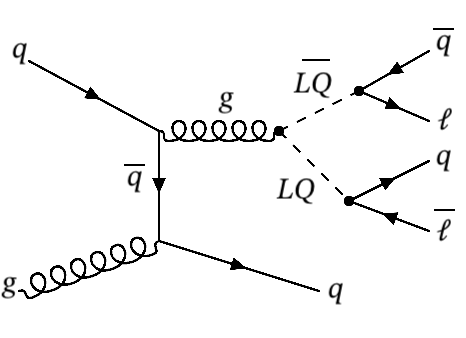}}\\\vspace{-0.5cm}
\subfloat[]{\includegraphics[scale=0.2]{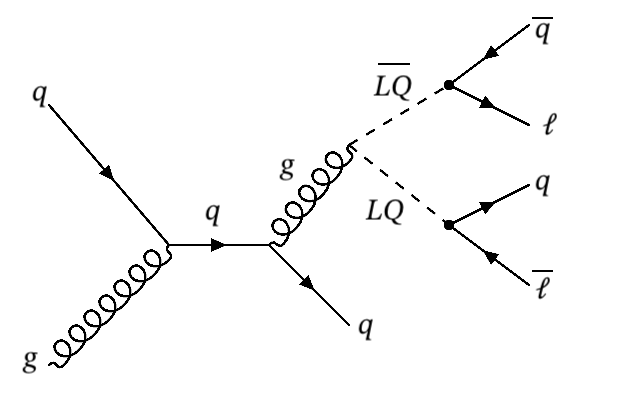}}
\subfloat[]{\includegraphics[scale=0.2]{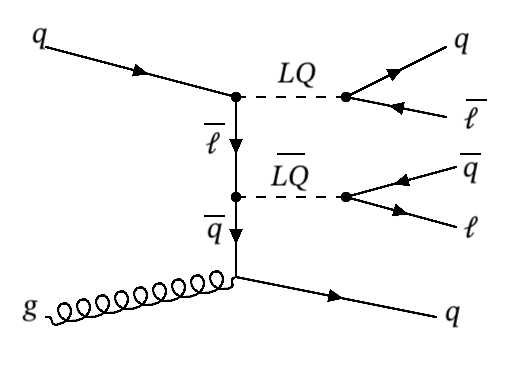}}
\caption{Examples of Feynman diagrams for leptoquarks originating via one of the three mechanisms highlighted in exclusive production: (i) the splitting of a gluon into two quarks, (ii) the emission of a gluon from a parton, and (iii) the exchange of a lepton. They are illustrated for t-channel (a-f), single (g-o), and pair (p-r) production of leptoquarks.}
\label{LQ_t_channel_single_pair}
\end{figure}

The t-channel process studied in literature refers to the diagram in Fig.~\ref{LQ_established_processes} (a) where the LQ interacts with two valence quarks in the initial state. However, such a production mode may be induced from the initial-state gluon splitting in either one or both incoming protons, as indicated in Fig.~\ref{LQ_t_channel_single_pair} (a, b). Another possibility is that a gluon or a lepton is exchanged, as illustrated in Fig.~\ref{LQ_t_channel_single_pair} (c, d), or a combination of the previous two cases, see Fig.~\ref{LQ_t_channel_single_pair} (e, f). 
Single LQ production is displayed on the diagrams in Fig.~\ref{LQ_established_processes} (b) where the LQ creation starts from a gluon and a quark. Cases of both initial gluon splitting or gluon emission from a parton are possible, as illustrated in Fig.~\ref{LQ_t_channel_single_pair} (g-m). Figure~\ref{LQ_t_channel_single_pair} (n, o) instead shows the case where a lepton is exchanged in the process. 
The quantum chromodynamic LQ pair production happens via a gluon that separates into two LQs, and the primary mode is shown in Fig.~\ref{LQ_established_processes} (c). Extra jets in final states coming from two LQ decays can originate from processes like those shown in Fig.~\ref{LQ_t_channel_single_pair} (p, q), while an example with a lepton exchange is displayed in Fig.~\ref{LQ_t_channel_single_pair} (r). 

The examples discussed above describe some of the possible modes to produce LQs with additional jets, and one can expect different variants allowed by crossing-symmetry of the particles involved in the interaction vertices. We note that combining the instances (1), (2), (3), as in diagram Fig~\ref{LQ_t_channel_single_pair} (e,f), further increases the number of jets and such diagrams are indeed produced by MadGraph, although their production cross section decreases as the number of jets increases.

\begin{figure}[!]
\subfloat[]{\includegraphics[scale=0.95]{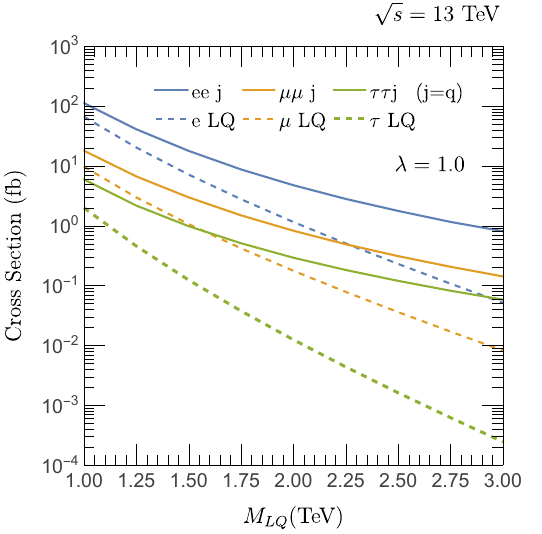}}
\subfloat[]{\includegraphics[scale=0.95]{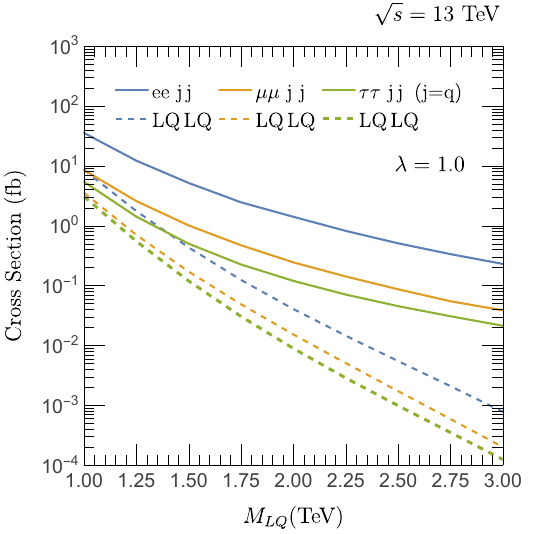}}\\
\subfloat[]{\includegraphics[scale=0.95]{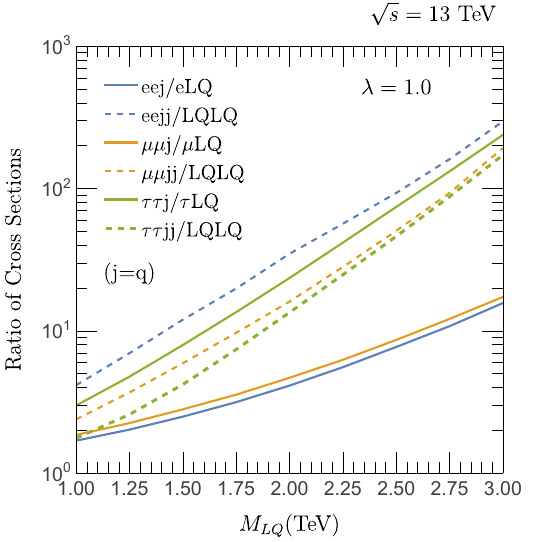}}
\caption{
Comparison of cross sections versus LQ mass for single (a) and pair (b) productions of Fig.\ref{LQ_established_processes} (b, c) with those obtained by specifying the corresponding dilepton plus jets (originating from quarks) final state, along with their ratio (c) for $\lambda_{eu}$, $\lambda_{{\mu}c}$, and $\lambda_{{\tau}b}$ set to 1.0.
}
\label{fig:xsec_sp}
\end{figure}

\begin{figure}[!h]
\centering
\includegraphics[width=0.65\textwidth]{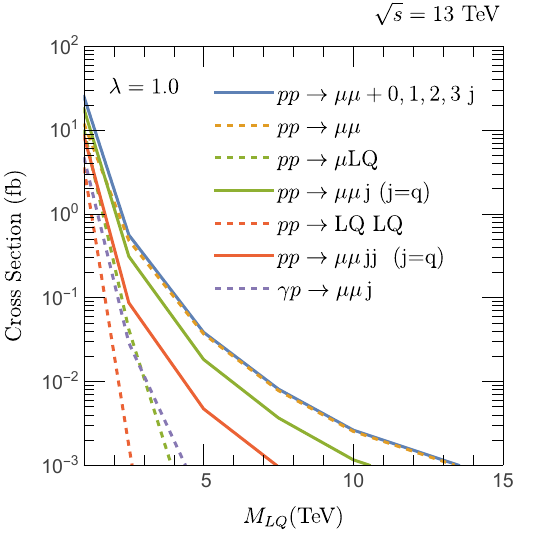}
\caption{Cross sections for different leptoquark production modes, for a coupling of the electron to the u quark $\lambda_{eu} =1$,  (a), and for a coupling of the muon to the c quark $\lambda_{\mu c} =1$, (b).}
\label{fig:xsec}
\end{figure}

In this study, we aim to understand the impact of these processes in addition to those usually considered in traditional LHC searches. {\color{black} To accomplish this, we compare the cross sections of the reference LQ production modes in Fig.\ref{LQ_established_processes} (b, c) 
with those obtained by specifying the corresponding dilepton plus jets (originating from quarks) final state 
that intercepts the new mechanisms described  above. The t-channel process in Fig.~\ref{LQ_established_processes} (a) is not considered, as it is produced with only two leptons in its traditional form. The comparison is presented in Fig.~\ref{fig:xsec_sp} with respect to both single (a) and pair (b) production, along with the ratio of the two production variants (c) for $\lambda_{eu}$, $\lambda_{{\mu}c}$, and $\lambda_{{\tau}b}$ equal to 1.0. These couplings encompass each fermion generation and are selected to represent all potential lepton-quark-LQ vertices. Indeed, the primary factor influencing the LQ production cross section, aside from its mass and coupling strength, is the specific quark to which it couples. Consequently, we can anticipate a comparable enhancement in the cross section for LQs coupling to other quarks, in alignment with proton density functions (PDFs), as well as for couplings involving neutrinos and cross-diagonal interactions between quarks and leptons.

Our analysis reveals a significant enhancement in the cross section across all cases when incorporating the new production modes. Specifically, there is an approximately 
70\%, 85\%, and threefold (fourfold, 2.4-fold, 75\%) increase against single (pair) production of Fig.~\ref{LQ_established_processes} for $\lambda_{eu}$, $\lambda_{{\mu}c}$, and $\lambda_{{\tau}b}$ set to 1 and an LQ mass of 1 TeV. This increase becomes larger at LQ mass of 3 TeV, corresponding to a factor of about 16, 17, 240 (300, 190, 170), respectively.

The observed enhancement is influenced by two factors: the flavor of the quark involved in the lepton-quark-LQ vertex according to the parton distribution functions of the proton and the diminishing resonant single and pair production cross section as a function of LQ mass due to the kinematic limit imposed by the LHC center-of-mass energy. The former implies that the new mechanisms contribute more for lower generation quarks with respect to pair production (Fig.~\ref{LQ_established_processes} (c)), which is mainly determined by gluon-gluon fusion and is independent of the quark generation, while they become more relevant for higher generation quarks against single production (Fig.~\ref{LQ_established_processes} (b)) that depends on one quark and is thus reduced for higher quark generations. The latter determines the larger increase at higher LQ masses, where the probability of creating a resonant LQ becomes lower and lower, while for the non-resonant mechanisms of Fig.~\ref{LQ_t_channel_single_pair} such a constraint is less stringent. For completeness, we should note that the new processes introduced here rely on the presence of gluons in the protons, which taper off at higher LQ masses, where valence quarks will dominate, as will the Drell-Yan t-channel. The gain of these new mechanisms for other beyond SM (or SM) particles will vary depending on their possible production modes.


This study marks the first presentation of such an analysis in the literature, demonstrating a significant benefit from considering the aforementioned production mechanisms. Based on these findings, we propose two distinct recommendations for future experimental strategies in searches for LQs or other particles at the LHC. First, we suggest revising the current approach where searches designed to detect events with a specific final state only consider particles from the corresponding primary production mode. As an example, it is customary that searches requiring two leptons and two jets in the final state target only LQ pair production (Fig.\ref{LQ_established_processes} (c)). Our study, instead, as indicated in Figs.~\ref{LQ_t_channel_single_pair} and \ref{fig:xsec_sp} for LQs, shows that the primary production modes contribute at varying jet multiplicities. Therefore, searches should be expanded to include all possible signal contributions, even if they target a single final state. Second, we remark the importance of considering t-channel contributions at any jet multiplicity, not just Fig.\ref{LQ_established_processes} (a), by exploiting its unique topology and using angular spectra of dileptons, which outperform traditional energy-related observables in sensitivity (see~\cite{CMS:2023qdw}). We will discuss in Section~\ref{search_strategy} how to leverage this feature to enhance the overall search sensitivity. 

As a final consideration, we address how to integrate all production modes to analyze the expected sensitivity to LQ signals at the LHC. Our focus is on an LQ with coupling $\lambda_{{\mu}c}$, as $\lambda_{eu}$ is predominantly determined by the Drell-Yan t-channel, and $\lambda_{{\tau}b}$ has a more complicated nature due to the presence of a $\tau$ lepton. Simply generating LQ events separately for each jet multiplicity and adding them would result in some degree of double counting of events. Instead, we produce exclusive LQ signal incorporating a lepton pair plus 0, 1, 2, and 3 jets (a jet being either a quark or a gluon to better simulate radiation effects), simultaneously. We correct for potential double counting issues using the jet matching technique detailed in section~\ref{signal_sample_generation}. The cross sections for all relevant processes are presented in Fig.~\ref{fig:xsec}, showing that exclusive LQ production dominates and is therefore utilized for our sensitivity analysis discussed below.
}

\subsection{Lepton initiated LQ production}
\label{lepton_initiated_LQ_production}
The idea that LQs may be produced from a lepton that stems via vacuum fluctuations in the proton was proposed a long time ago~\cite{Ohnemus:1994xf}. It has become recently feasible due to an improved estimate of the lepton density functions (LDFs)~\cite{Buonocore:2020erb, Buonocore:2022msy, Korajac:2023xtv} and it has been used at CMS~\cite{CMS:2023bdh}. Here, we discuss how such a process enters our view to proceed to a global search for LQs. 

The first point to remark on is that the works~\cite{Buonocore:2020erb, Buonocore:2022msy, Korajac:2023xtv, CMS:2023bdh} study s-channel LQ production only. However, t-channel LQ exchange is possible and should be considered. The second aspect to note is that one can consider lepton-induced processes using the exclusive production approach introduced here by noting that all the processes illustrated in Fig.~\ref{LQ_t_channel_single_pair} and discussed above are equally possible by simply substituting a gluon that splits into two quarks with a photon that splits into two leptons. 

\begin{figure}[!h]
\centering
\subfloat[]{\includegraphics[scale=0.35]{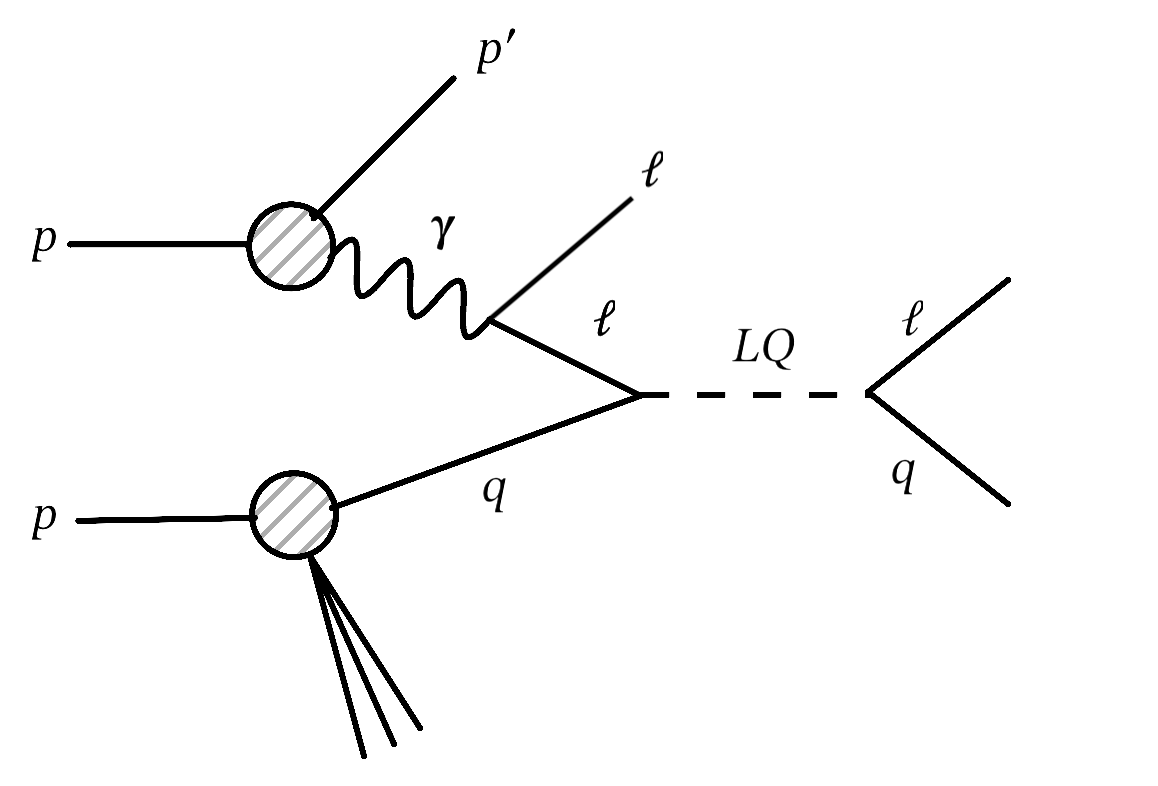} }
\subfloat[]{\includegraphics[scale=0.35]{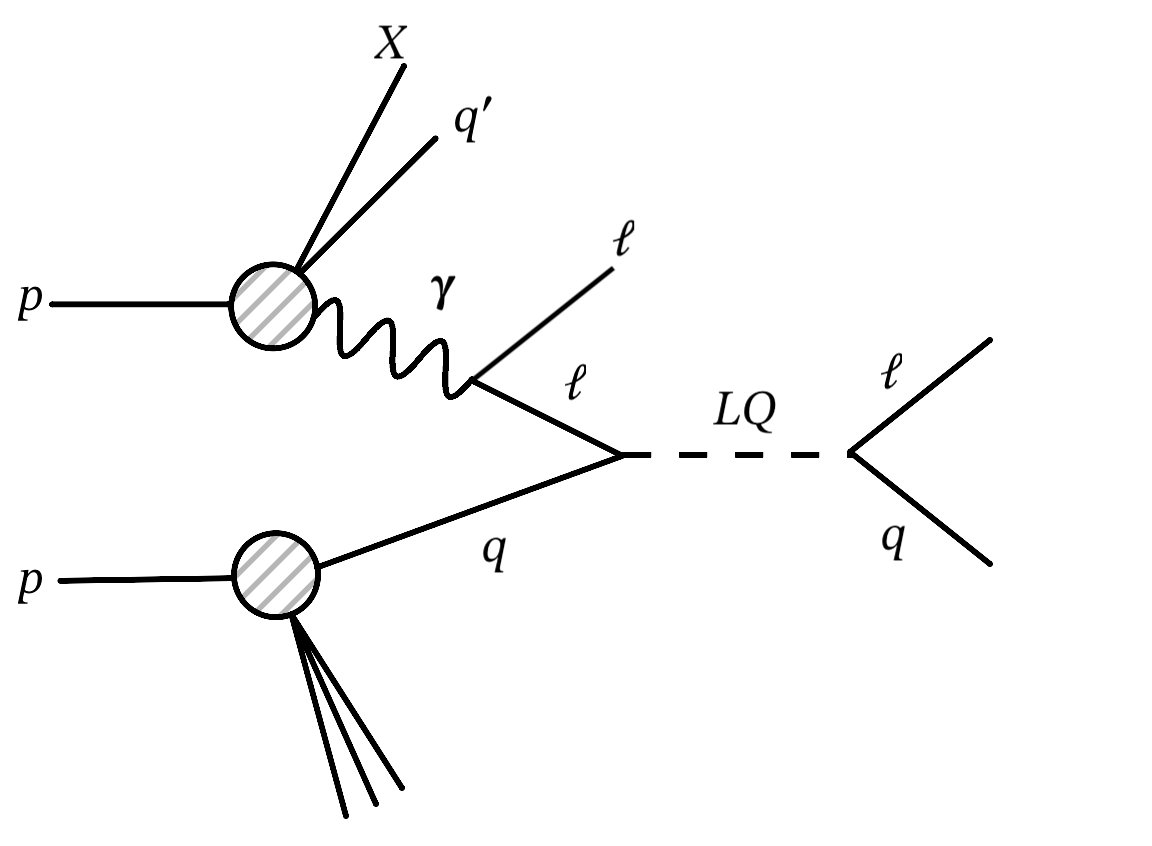} } \\
\caption{Feynman diagrams for the single leptoquark production for the cases where a photon that splits into two leptons is radiated by the proton elastically (a) or inelastically (b). 
}
\label{lepton_induced}
\end{figure}

There is another more general consideration to mention in relation to the lepton that produces the LQ, and it concerns whether its mother particle, the photon, for example, stems from elastic (the proton remains intact, Fig.~\ref{lepton_induced} (a)) or inelastic (the proton does not remain intact, Fig.~\ref{lepton_induced} (b)). 
For the elastic case, we define one of the protons to be a photon, while 
for the inelastic case, we include the photon in the proton without changing the default type. 
For $\lambda_{\mu c}$ = 1, we see that the ratio of the contribution to the cross section from the elastic to the inelastic case is about 27\%. Despite the lower ratio for the elastic case, we point out that a dedicated search relying on the elastic case would have an enormous advantage in mitigating the SM background, as pointed out in several investigations~\cite{TOTEM:2021zxa}. 

By comparing the total cross section obtained summing up both elastic and inelastic cases
with that using the LDFs as in Fig.\ref{LQ_established_processes} (d), we see that the two are consistent within 20\%. 

\section{Search strategy}
\label{search_strategy}
\begin{figure}[!h]%
    \centering
    \subfloat[]{{\includegraphics[width=5.75cm]{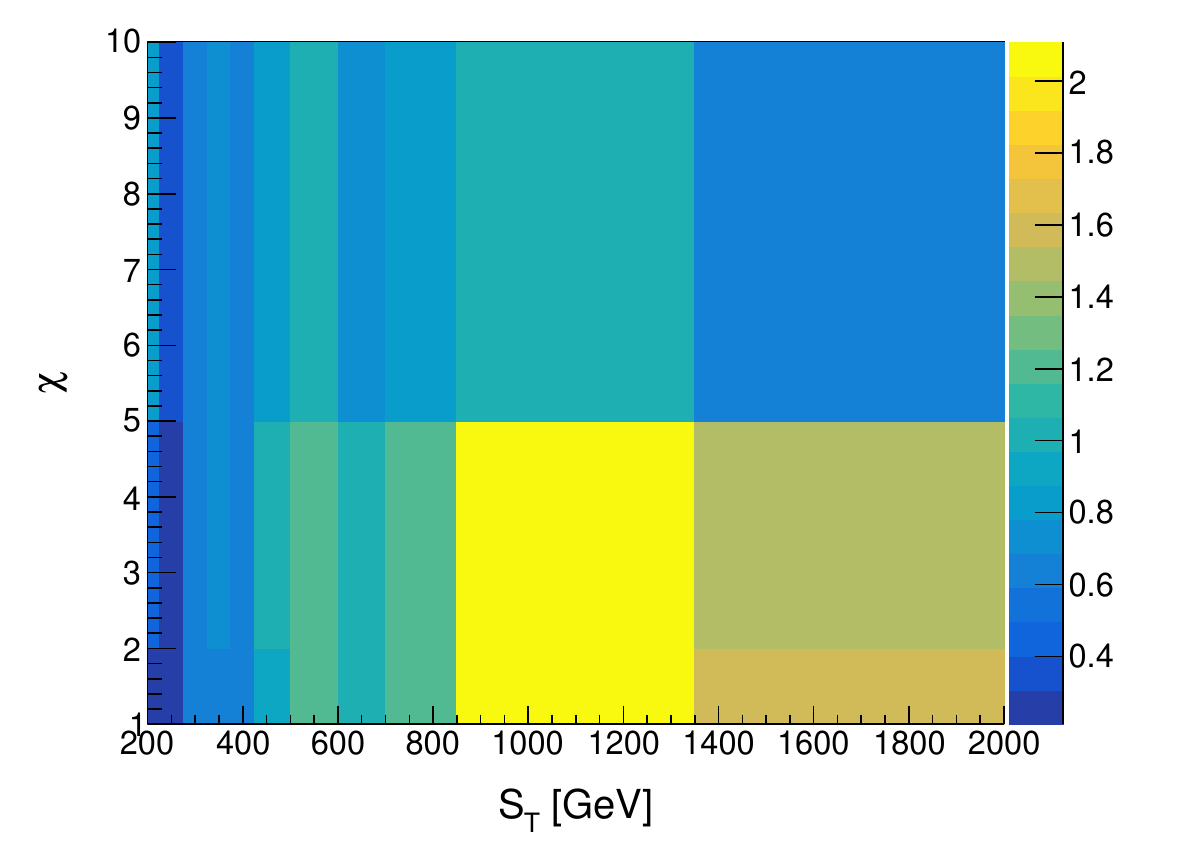} }}%
    \subfloat[]{{\includegraphics[width=5.75cm]{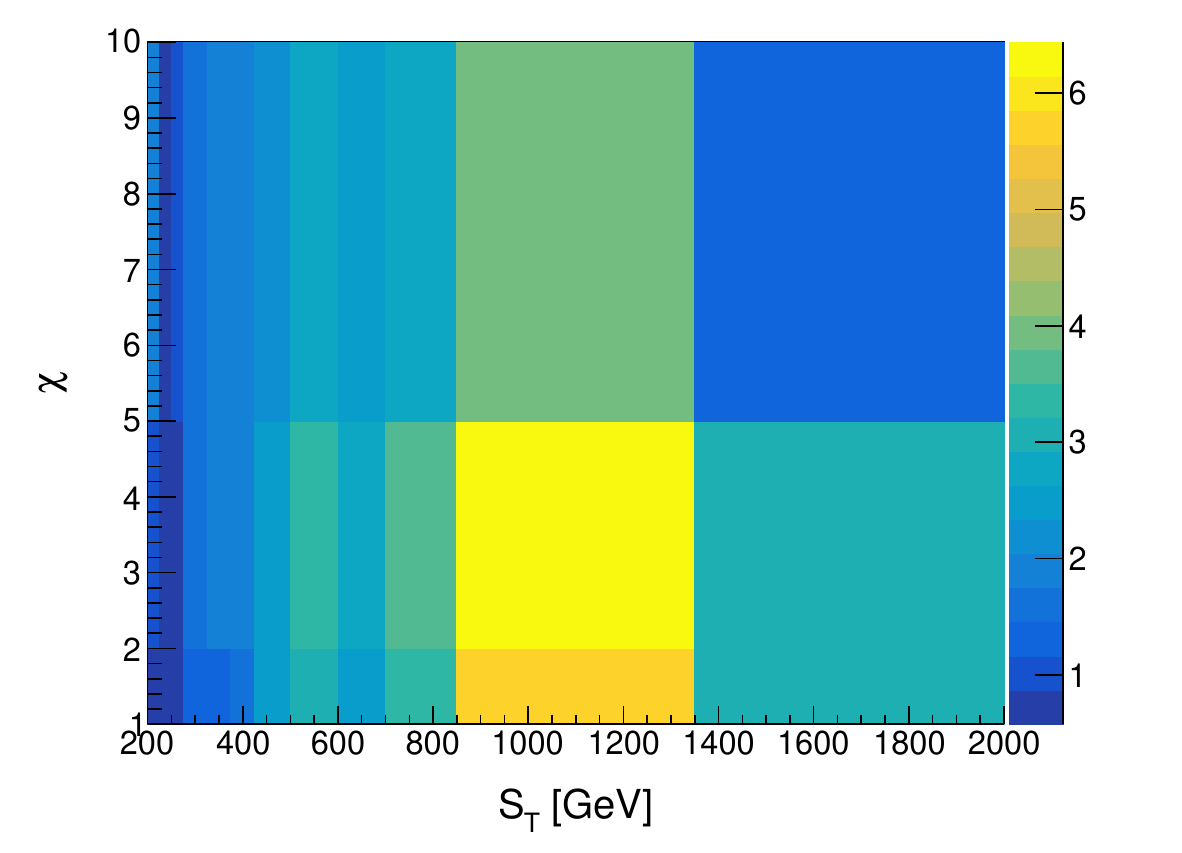} }}%
    \subfloat[]{{\includegraphics[width=5.75cm]{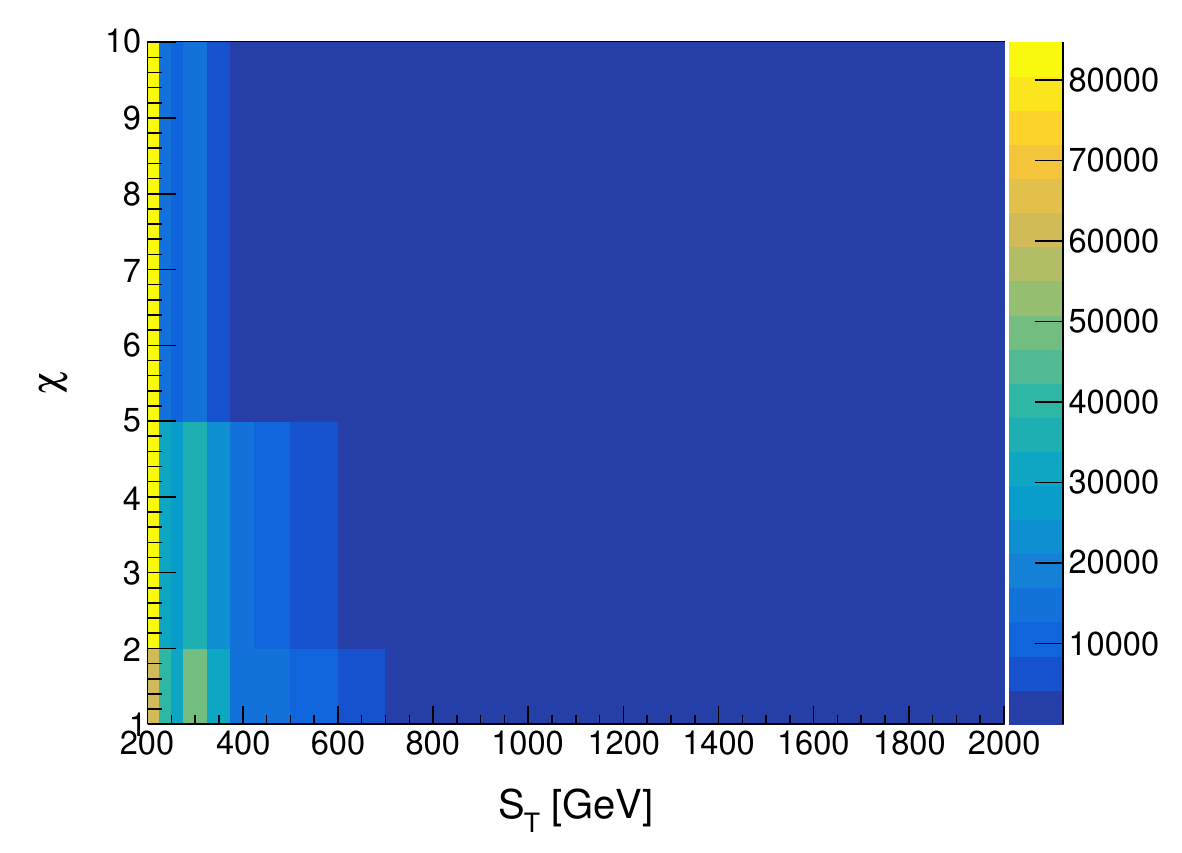} }}%
    \caption{Distributions of the pp events passing the baseline selection specified in the text and requiring two jets in the two-dimensional plane of angular $\chi$ and energy $\text{S}_{\text{T}}$ observables defined in the paper. The plots are shown for two signals hypotheses with leptoquark mass equal to 2 (4) TeV and $\lambda_{\mu c}$ coupling of 1.0 (2.5) on the left (center) and the SM background on the right.}%
    \label{fig:2Ddistrib}%
\end{figure}

The signal we look for produces a signature with two leptons in the final state plus a given number of jets (0, 1, 2, or at least 3). The samples described in Section~\ref{signal_sample_generation} for the $\lambda_{\mu c}$ coupling, as discussed in Section~\ref{exclusive_lq_production}, are used in the event selection.  
The events pass the signal selection if they satisfy the following baseline requirements: exactly two muons with transverse momentum $p_{\rm T}>$ 20 GeV, pseudorapidity $|\eta|<$ 2.5, and invariant mass greater than 120 GeV; 
missing transverse energy lower than 50 GeV; and no jets originating from b quarks. 
We find that the selection efficiency is about 30\% or higher from masses of 1 TeV and all coupling values. 
\\
Jets are chosen to have $p_{\rm T}>$ 20 GeV and $|\eta|<$ 5, and to be spatially separated from the selected muons by $\Delta R >$ 0.4, where $\Delta R = \sqrt{\smash[b]{(\Delta\eta)^2+(\Delta\phi)^2}}$ and $\phi$ is the azimuthal angle. We do not set any condition on a minimal number of jets, as we separate the signal region events into categories with 0, 1, 2, or at least 3 jets. 
\\
{\color{black} As discussed in Section~\ref{exclusive_lq_production}, one significant finding of this study is the contribution from t-channel processes among the new mechanisms illustrated in Fig.~\ref{LQ_t_channel_single_pair}, in addition to the Drell-Yan production depicted in Fig.~\ref{LQ_established_processes}. To optimize our search criteria to be sensitive to these processes, we consider angular spectra of dileptons. This method offers superior discrimination against background (see~\cite{CMS:2023qdw}) compared to methods considering the energy-related observables typically employed in LQ searches, which are instead well-suited for signal events featuring resonant LQs. Given our goal of building a comprehensive search strategy, we propose leveraging both these aspects simultaneously to enhance our sensitivity to the LQ signal.
} 
In Fig.~\ref{fig:2Ddistrib} we show the distribution from pp events from processes including two jets in the two-dimensional plane of $\chi$ and $\text{S}_{\text{T}}$, where $\chi = e^{|\eta_1-\eta_2|}$ and $\eta_1$, $\eta_2$ are the pseudorapidities of the two muons, while $\text{S}_{\text{T}}$ is defined as the sum of the scalar $\text{p}_{\text{T}}$ of the leptons and selected jets. The plots are shown for two signals hypotheses, with LQ mass equal to 2 (4) TeV and $\lambda_{\mu c}$ coupling of 1.0 (2.5) on the left (center) and the SM background on the right.

Background events are generated at leading order (LO) with PDF NNPDF31-lo-as-0118 \cite{NNPDF:2017mvq}, and scaled using next-to-leading-order (NLO) cross sections where appropriate. The processes we consider are W + jets ($W\rightarrow l+\nu$), Drell-Yan (charged lepton decays) + jets, single top, $t\bar{t}$, and diboson (WW, WZ, and ZZ). Background processes are generated with sufficiently high jet multiplicities to yield an accurate comparison against our targeted signal. Generator-level production is then interfaced with Pythia 8.2 \cite{Sjostrand:2014zea} for parton showering and hadronization using the Monash 2013 tune \cite{Skands:2014pea}. Detector response simulation is obtained via the Delphes framework~\cite{deFavereau:2013fsa}.  As in our signal events, jet matching is implemented using the MLM algorithm to avoid double counting of events after showering and hadronization.

%
%
%




\section{Expected sensitivity at LHC and HL-LHC}
\label{expected_sensitivity}
\begin{figure}[!h]
    \centering
    \subfloat[]{{\includegraphics[width=8.2cm]{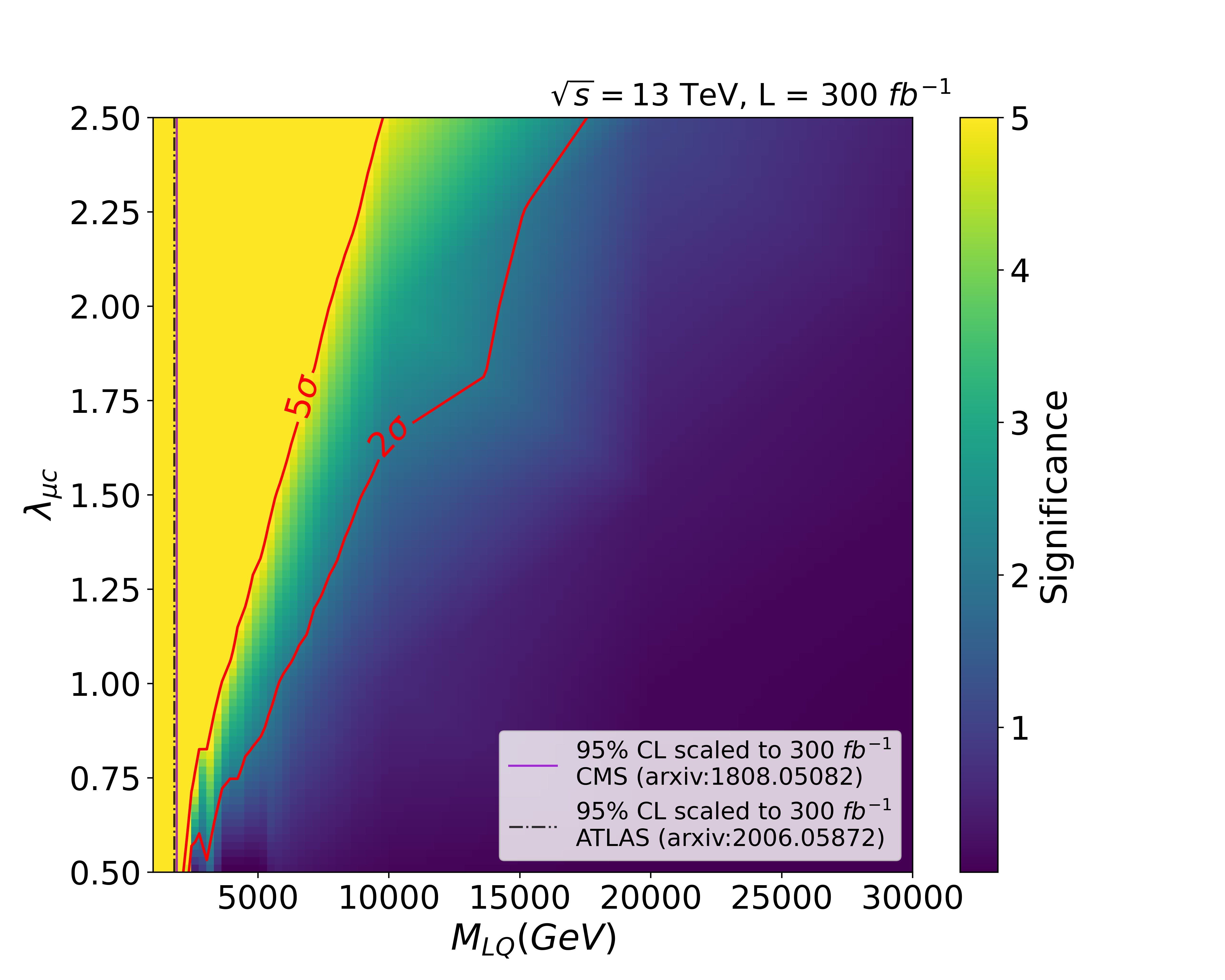}}}%
    \subfloat[]{{\includegraphics[width=8cm]{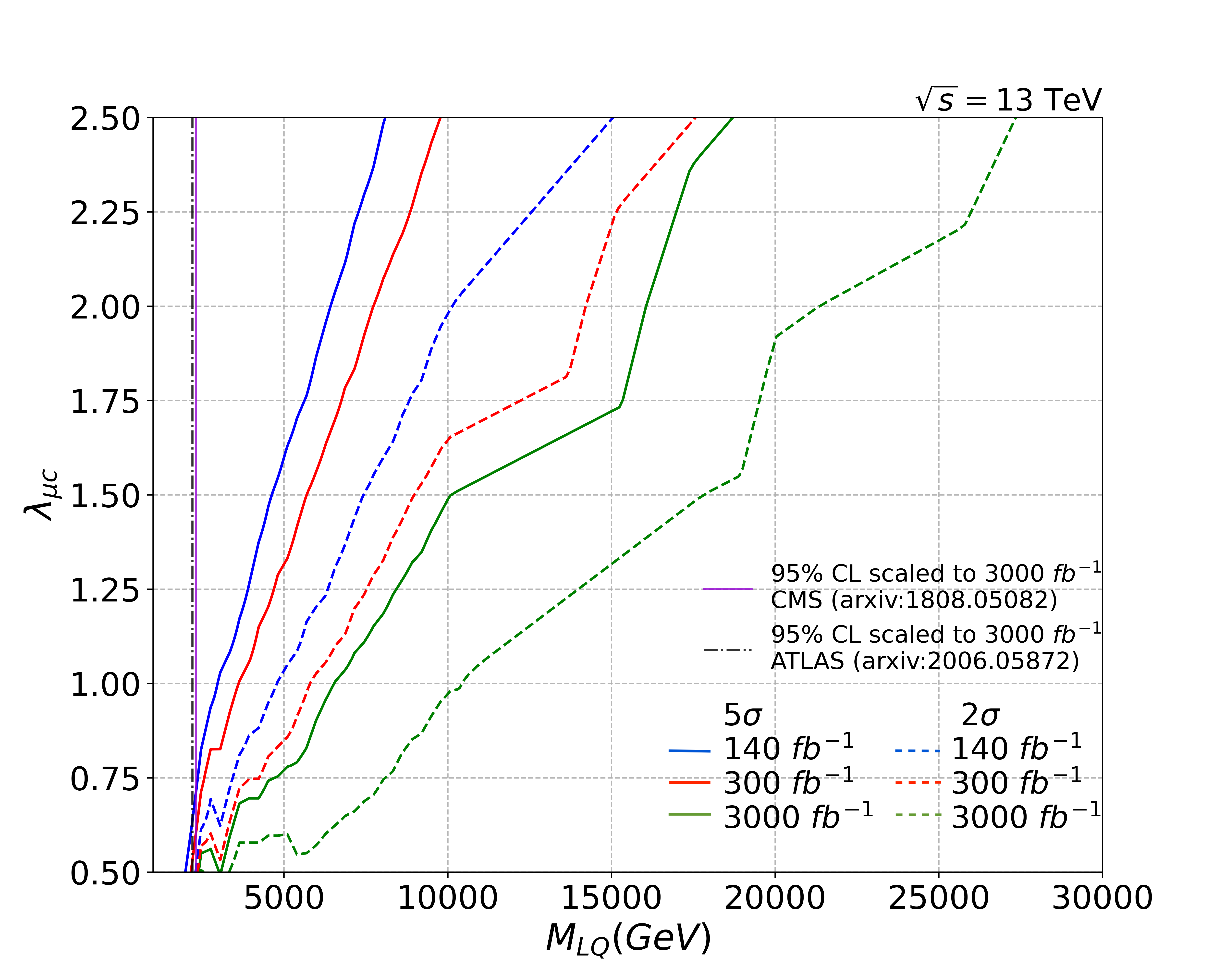}}}%
 \caption{Signal significance in the plane of the coupling $\lambda_{\mu c}$ and the mass of the LQ using $pp$ and $\gamma p$ events at $\sqrt{s}= 13$ TeV expected for 300 $fb^{-1}$ (left) and different luminosity scenarios (right). The colored solid and dashed lines represent 5 and 2 $\sigma$ levels for the analysis presented in this text, while the vertical lines show the most recent exclusion limits results at 95\% confidence level from a search for LQs from ATLAS~\cite{ATLAS:2020dsk} (black dashed) and CMS~\cite{CMS:2018lab} (violet plain) collaborations scaled for 300 (left) and 3000 (right) fb$^{-1}$.} 
 \label{significance} 
\end{figure}

The distributions of the $\chi$ and $\text{S}_{\text{T}}$ variable per jet multiplicity for the events of the signal regions are used as the final discriminant between the LQ signal and the background. We sum up signal events from both the $pp$ and $\gamma p$ collisions, accounting for production cross section, efficiency, and luminosity, and use a profile-binned likelihood statistical test \cite{pdg_stat} for assessing the expected signal significance. 
Nuisance parameters encompassing systematic uncertainties are introduced for the signal (background), incorporating log-normal priors. We follow the same prescriptions used in the published LQ searches~\cite{CMS:2018lab, CMS:2018ncu} and we briefly summarize them below. The uncertainties arising from the selection of the parton distribution function in the MC production is $2.8 \%$ ($3.0 \%$), while that related to additional interaction vertices in the event amount to $0.2 \%$ ($1.0 \%$). The uncertainties associated with jet and lepton energy resolution are $0.1 \%$ ($1.7 \%$) and $0.2 \%$ ($5.3 \%$), respectively. Additionally, jet and lepton energy scale uncertainties are included with values of $0.5 \%$ ($0.9 \%$) and $1.5 \%$ ($2.5 \%$). The uncertainties for leptonic reconstruction and identification efficiency stand at $3.0 \%$ and $1.3 \%$ ($0.3 \%$). 
For the calculation of expected significance the ``Combine Tool'' is used \cite{combine}.
In Fig.~\ref{significance}, we present contour plots of the expected significance in the two-dimensional plane of the coupling $\lambda$ and the mass of the LQ. These plots are obtained at the center-of-mass energy of $13$ TeV and an integrated luminosity of 300 fb$^{-1}$ (left), as a possible value of the luminosity expected by the end of the ongoing data-taking period at the LHC referred to as Run 3, and varying luminosity assumptions (right) at 140 (blue), 300 (red), 3000 (green) fb$^{-1}$, for the 5 $\sigma$ (plain line) and 2 $\sigma$ (dashed line) significance levels.  The vertical lines show the most recent exclusion limits results at 95\% confidence level from a search for LQs from ATLAS~\cite{ATLAS:2020dsk} (black dashed) and CMS~\cite{CMS:2018lab} (violet plain) collaborations scaled for 300 (left) and 3000 (right) fb$^{-1}$. The results obtained with the exclusive LQ approach report a 5 (2) $\sigma$ significance for LQ masses up to 
2.0, 2.2, 2.5 (2.3, 2.4, 3.4) TeV, for $\lambda_{\mu c}$ = 0.5, and 8.0, 9.7, 18.6 (14.9, 17.5, 27.3) TeV, for $\lambda_{\mu c}$ = 2.5, at 140, 300, 3000 fb$^{-1}$. 
While ATLAS and CMS scaled results would exclude a LQ coupling to a muon - \textit{c} quark pair up to masses of 1.7, 1.8, 2.2, and 1.8, 1.9, 2.3 TeV regardless of the coupling value and for the same luminosity scenarios. With the strategy developed in this article, we obtain that the 2 $\sigma$ significance with 140 fb$^{-1}$ is comparable to the scaled upper limit on the LQ mass expected by ATLAS and CMS with 3000 fb$^{-1}$, for the lowest coupling value that we consider.
   
\section{Summary and remarks}
\label{summary_remarks}
As the LHC proceeds in collecting more data at the utmost center-of-mass energy and with the slow relative increase in luminosity anticipated by the High-Luminosity initiative of the accelerator, the absence of compelling evidence for new physics prompts a critical consideration. It becomes increasingly important to assess whether existing searches have thoroughly investigated all conceivable paths or whether new approaches may still be formulated to enhance the discovery potential and to unlock access to a phase space that was not considered reachable before. 

In this paper, we have tackled this question by conducting a comprehensive study of the well-motivated, known, and extensively researched LQ particle. Although we carry out a study that is valid for any model, we have relied on a novel scenario of LQ composite particles via the NJL model with effective strong four-fermion couplings~\cite{Xue2017,Xue2023}. 
{\color{black} This is an important addition to the literature, as in the event of discovering a beyond Standard Model particle consistent with an LQ, it will be crucial to ascertain its nature and determine its classification within the various discernible models. In particular, the NJL model} 
introduces a different concept of compositeness than those considered at LHC~\cite{ATLAS:2023kek, ATLAS:2019zfh, ATLAS:2017dpx, CMS:2018nlk, CMS:2020cay, CMS:2022chw, CMS:2018sfq, CMS:2015lgt} and offers a broad variety of new composite particles: integer charge bosons $\Pi^{f}\sim (\bar\psi^{f}_{_R} \psi^{f}_{_L})$ and fermions $F^{f}_R\sim \psi^{f}_{_R} (\bar\psi^{f}_{_R} \psi^{f}_{_L})\sim \psi^{f}_{_R}\Pi^{f}$~\cite{Leonardi:2018jzn}, fractional charge bosons such as the LQ bosons studied here, composite LQ fermions, and composite Higgs to be studied in future works. These could manifest in signatures that have not yet been explored at the LHC. 

Compared to existing investigations that rely on the primary LQ production modes in Fig.~\ref{LQ_established_processes}, for the first time, we have examined the impact on the sensitivity of the search for particles when they are produced in association with additional jets. We have found that several new mechanisms are possible besides the classical pair, single, and Drell-Yan non-resonant LQ generation, as discussed in Section~\ref{exclusive_lq_production} and illustrated in Fig.~\ref{LQ_t_channel_single_pair}. We refer to the ensemble of all allowed processes as ``exclusive'' LQ production, in analogy to the terminology adopted in inelastic scattering, where all reaction products are involved. 

{\color{black} We have observed that the exclusive cross section exceeds those of the mechanisms considered in traditional searches. These carry an increase of approximately 70\%, 85\%, and threefold (fourfold, 2.4-fold, 75\%) over the single (pair) production modes shown in Fig.~\ref{LQ_established_processes} respectively for $\lambda_{eu}$, $\lambda_{{\mu}c}$, and $\lambda_{{\tau}b}$ with their value set to 1 and an LQ mass of 1 TeV. This increase becomes larger at an LQ mass of 3 TeV, corresponding to a factor of about 16, 17, 240 (300, 190, 170), respectively for the considered couplings over single (pair) production. This result represents an outstanding gain in discovery potential due to the enhanced signal cross section. Additionally, the signal significance is further increased because, for the same primary mode, the exclusive production captures new mechanisms at higher jet multiplicities, which have been shown to further mitigate the SM background~\cite{CMS:2023byi, ATLAS:2021mla, ATLAS:2019odt, CMS:2019mij, Abdullah:2018ets, Abdullah:2017oqj, Barbosa:2022mmw, Abdullah:2019dpu}. 

Given all the findings presented in this paper, we propose two fundamental recommendations for developing strategies to investigate new physics, whether for LQs or other particles. The first is the inclusion of all possible signal contributions across varying jet multiplicities, employing the exclusive approach rather than relying solely on the primary production modes, even if an analysis targets a single final state. The second is to use angular spectra in conjunction with energy-related observables to fully exploit t-channel contribution and achieve the highest significance. 
}


Moreover, we have expanded this approach referring to LDFs and we also pursued an unprecedented distinction between inelastic and elastic modes, the former resulting in about 80\% of the total although the latter would benefit from reduced background contamination~\cite{TOTEM:2021zxa}. 

{\color{black}Finally, we propose a global strategy as our key recommendation, to intercept t-channel, single, and pair signal generation at different jet multiplicities and simultaneously using the $\text{S}_{\text{T}}$ and $\chi$ distributions.}
We anticipate that an optimal combination of the energetic and angular variables could be exploited by relying on machine learning techniques. As shown in Fig.~\ref{significance}, we achieve a significance of 5 (2) $\sigma$ for LQ masses up to  
2.0, 2.2, 2.5 (2.3, 2.4, 3.4) TeV, for $\lambda_{\mu c}$ = 0.5, and 8.0, 9.7, 18.6 (14.9, 17.5, 27.3) TeV, for $\lambda_{\mu c}$ = 2.5, at 140, 300, 3000 fb$^{-1}$.
The significance of 2 $\sigma$ with 140 fb$^{-1}$ is comparable to the scaled upper limit on the LQ mass expected by ATLAS~\cite{ATLAS:2020dsk} and CMS~\cite{CMS:2018lab} with 3000 fb$^{-1}$ for the lowest coupling value that we consider. 

These results are notably relevant for BSM scenarios with preferential couplings to higher fermion generations, including models that have the potential to explain some of the persistent, yet to be confirmed, anomalies in high energy physics (such as the aforementioned $B\rightarrow D^*$ and $(g-2)_\mu$ precision measurements), as they allow to probe a much wider region of the LQ coupling-mass plane compared to traditional searches and especially for couplings to higher generation fermions. 

{\color{black}In light of the discovery potential and the results presented in this paper, we assert that an exclusive approach to LQ searches should be incorporated into future investigations at the LHC. In generalizing these results, we find that the exclusive paradigm and the aforementioned recommendations should be applied to the exploration of any new particles.}


%
%

\begin{acknowledgments}
This work has received funding from the INFN project ENP (Exploring New Physics). J. Gaglione, A. Gurrola, and F. Romeo acknowledge the funding received from the Physics and Astronomy department at Vanderbilt University and the US National Science Foundation. This work is supported in part by NSF Award PHY-1945366 and a Vanderbilt Seeding Success Grant. The work of M. Presilla is supported by the Alexander von Humboldt-Stiftung. H. Sun is supported by the National Natural Science Foundation of China (Grant No.12075043).
\end{acknowledgments}

\bibliographystyle{unsrt}
\bibliography{LQ-Composite-Boson.bib}

\end{document}